\documentclass[12pt]{article}
\usepackage{amsmath}
\usepackage{amssymb}
\usepackage{fullpage}
\usepackage[dvips]{epsfig}

\def\##1{{\bf #1}}
\def\=#1{\underline{\underline #1}}

\def\le{\left(}
\def\ri{\right)}
\def\les{\left[}
\def\ris{\right]}
\def\lec{\left\{}
\def\ric{\right\}}

\def\c#1{\cite{#1}}

\def\r#1{(\ref{#1})}

\def\epso{\epsilon_0}
\def\muo{\mu_0}

\def\epsr{\epsilon_r}
\def\epsrr{\epsilon_r'}
\def\epsri{\epsilon_r''}
\def\mur{\mu_r}
\def\murr{\mu_r'}
\def\muri{\mu_r''}
\def\nr{n}
\def\nrr{n'}
\def\nri{n''}
\def\om{(\lambda_0)}

\def\rt{({\bf r},t)}
\def\zt{(z,t)}

\def\ux{\hat{\#x}}
\def\uy{\hat{\#y}}
\def\uz{\hat{\#z}}

\begin{document}

\noindent{\large\bf On reflection from a half-space with
negative real permittivity and permeability: Time-domain
and frequency-domain results}

\vskip 1.0 cm

\noindent {\bf Jianwei Wang}$^1$ and {\bf Akhlesh Lakhtakia}$^1$
\vskip 0.4 cm

\noindent $^1$CATMAS --- Computational \& Theoretical Materials Sciences Group\\
\noindent Department of Engineering Science and Mechanics\\
\noindent Pennsylvania State University\\
\noindent University Park, PA 16802--6812
\vskip 1 cm

\noindent {\bf ABSTRACT:}
The  reflection of a normally incident wideband pulse by a 
half--space whose permittivity and permeability
obey the one--resonance Lorentz model is calculated. The results
are
compared with those from frequency--domain reflection analysis.
In the spectral regime wherein the real parts of the permittivity
and the permeability are negative, as well as in  small
adjacent neighborhoods of that regime, the time--domain
analysis validates the conclusion that it is sufficient
to ensure that the imaginary
part of the refractive index is  positive~---~regardless
of the sign of the real part of the refractive index.


\section{Introduction}
Take a half--space occupied by  an isotropic, homogeneous,
dielectric medium
of relative permittivity $\epsr\om =\epsrr\om +i\epsri\om$, where 
$\lambda_0$ is the free--space wavelength. When a plane wave with 
 electric field
phasor of unit amplitude is
incident normally on this half--space, the amplitude $r\om$
of the
electric field phasor of the reflected plane wave is given
by
\begin{equation}
\left.
\begin{array}{l}
r\om = \frac{\nr\om-1}{\nr\om+1}\\[10pt]
\nr\om = \sqrt{\epsr\om}
\end{array}\right\}\,,
\end{equation}
where $\nr\om$
is the refractive index. The foregoing standard
result is  found in countless many textbooks; see,
e.g., Born \& Wolf \c{BW} and Jackson \c{J}.
It can be applied to insulators, poor conductors
as well as good conductors.

Let the medium occupying the half--space also have magnetic properties
represented by the relative permeability $\mur\om=\murr\om+i\muri\om$.
Then, the previous result has to be modified as follows:
\begin{equation}
\label{rFD}
\left.
\begin{array}{l}
r\om = \frac{\les\nr\om/\mur\om\ris-1}{\les\nr\om/\mur\om\ris+1}\\[10pt]
\nr\om = \sqrt{\epsr\om\,\mur\om}
\end{array}\right\}\,.
\end{equation}

This result \r{rFD} for the
reflection coefficient $r\om$ gains particular prominence in the
context of composite materials that supposedly
exhibit $\lec\epsrr<0, \, \murr<0\ric$  and
$\lec\epsri\geq 0, \, \muri\geq 0\ric$ 
in some spectral regime \c{SSS}~---~the time--dependence
$\exp(-i\omega t)$ being implicit for all electromagnetic fields,
with $\omega$ as the angular frequency.
Shelby {\em et al.\/} \c{SSS}
 argued that the real part  of
$\nr\om =\nrr\om+i\nri\om$ must then be chosen to be negative.
Consequently, it was concluded \c{SK} that ``[p]lane
waves appear to propagate from plus and minus infinity
towards the source ...... [but] ...... clearly energy propagates
outwards from the source."
The conclusion
has been employed to deduce the amplification of evanescent
plane waves, leading to a proposal for aberration--free lenses \c{P}.

Certainly, energy does not flow in the
{\em backward\/} direction. Rather, the forward direction
is defined 
 as the direction of the energy transport, while
the phase velocity
{\em may} then be pointed in the backward direction~---~a known phenomenon \c{IVL}.
Assuming the one--resonance
Drude model for both $\epsr\om$
and $\mur\om$, confining themselves to the
spectral  regime wherein $\epsrr\om <0$
and $\murr\om < 0$, and impedance--matching the
dielectric--magnetic half--space to
free space, Ziolkowski and Heyman \c{ZH}  pointed out that
 the reflection
coefficient $r\om$ must have finite magnitude. Employing
the one--resonance Lorentz model for both $\epsr\om$ and $\mur\om$,
McCall {\em et al.\/} \c{MLW} recently deduced that 
\begin{itemize}
\item[(i)] both $\epsrr\om$
and $\murr\om$ need not be negative for $\nrr\om$ to be negative
and $\nri\om$ to be positive, and 
\item[(ii)] the magnitude
of the reflection coefficient
$r\om$ does not then exceed unity.
\end{itemize}
Let us emphasize that
the spectral regime for $\nrr\om<0$ overspills
into the neighborhoods of the regime wherein both
$\epsrr\om$ and $\murr\om$ are negative.

Whereas frequency--domain analysis for the response of
a half--space calls for the
determination of the signs of $\nrr\om$
and $\nri\om$, the refractive index does
not enter time--domain analysis.
Therefore, time--domain analysis offers an independent way
of assessing the results of frequency--domain analysis. In
this communication, we present our conclusions from the response
of a dielectric--magnetic half--space with negative
real refractive index in a certain
spectral regime to a normally incident wideband pulse. The Fourier
transform of the reflected pulse confirms the frequency--domain
calculations of
$r\om$ in that spectral regime, and
underscores
 the requirement of $\nri\om >0$ for real materials
(which cannot be non--dissipative due to causality) to be sufficient.

\section{Theory in Brief}
Consider the region $\left\{z\geq 0,\,t\geq 0\right\}$ of 
spatiotemporal space. The half--space $z\geq z_\ell$
is occupied by a homogeneous
dielectric--magnetic medium
whose constitutive relations are given by
\begin{equation}
\left.\begin{array}{l}
{\bf D}\rt = \epso\les{\bf E}\rt+ (\chi_e\ast{\bf E})\rt\ris\\
{\bf B}\rt =\muo\les{\bf H}\rt + (\chi_m\ast{\bf H})\rt\ris
\end{array}\right\}\,,\quad z\geq z_\ell\,,
\end{equation}
where the asterisk denotes the convolution operation \c{Good} with respect
to time, while the susceptibility functions
\begin{eqnarray}
\nonumber
\chi_{e,m}(t) &=& p_{e,m}\, \frac{2\pi c_0}{\lambda_{e,m}}\,
\sin\le\frac{2\pi c_0 t}{\lambda_{e,m}}\ri\,\\
&&\quad\qquad\times
\exp\le-\frac{c_0 t}{M_{e,m}\lambda_{e,m}}\ri \,{\cal U}(t)
\,
\end{eqnarray}
obey the one--resonance Lorentz model. These susceptibility
functions correspond
to 
\begin{eqnarray}
\label{eFD}
\epsr\om &=& 1 + \frac{p_e}{1+(N_e^{-1} - i\lambda_e\lambda_0^{-1})^2}\,,\\
\label{mFD}
\mur\om &=& 1 + \frac{p_m}{1+(N_m^{-1} - i\lambda_m\lambda_0^{-1})^2}\,
\end{eqnarray}
in the frequency
domain \c{BH}.
In these expressions, $\epso$ and $\muo$ are
the permittivity and the permeability of free space;
 $c_0 =(\epso\muo)^{-1/2}$ is the speed of light in free space;
${\cal U}(t)$ is the unit step function;
$p_{e,m}$ denote the so--called oscillator strengths; while
$N_{e,m}=2\pi M_{e,m}$ and $\lambda_{e,m}$ determine
the resonance wavelengths and the linewidths.
The region $z\leq z_\ell$ is vacuous.

At time $t=0$, an amplitude--modulated wave is supposedly launched
normally from the plane $z=0$; therefore, all fields
are independent of $x$ and $y$. The initial and boundary conditions
on ${\bf E}\zt$ and ${\bf H}\zt$ are as follows:
\begin{equation}
\left.\begin{array}{l}
{\bf E}(z,0) = {\bf 0} \quad \forall z\geq 0\\[10pt]
{\bf H}(z,0) = {\bf 0} \quad \forall z\geq 0\\[10pt]
{\bf E}(0,t) = g(t) \sin\le\frac{2\pi c_0 t}{\lambda_{car}}\ri 
\,\uy \quad \forall t\geq 0\\[10pt]
{\bf H}(0,t) = - (\epso/\muo)^{1/2}\,
g(t) \sin\le\frac{2\pi c_0 t}{\lambda_{car}}\ri 
\,\ux \quad \forall t\geq 0
\end{array}\right\}\,.
\end{equation}
Whereas $\lambda_{car}$ is the carrier
wavelength, the function
\begin{equation}
g(t) = \frac{c_0t}{2\lambda_{car}}\,\exp(-c_0t/\lambda_{car})\,
\end{equation}
was chosen to represent the pulse.
The cartesian unit vectors are denoted by $(\ux,\,\uy,\,\uz)$.

The differential equations to be solved
are as follows:
\begin{eqnarray}
\frac{\partial}{\partial z} E_y\zt
=\muo \les \frac{\partial}{\partial t} H_x\zt
+\frac{\partial}{\partial t} (\chi_m\ast H_x)\zt\ris\,,\\
\frac{\partial}{\partial z} H_x\zt
=\epso \les \frac{\partial}{\partial t} E_y\zt
+\frac{\partial}{\partial t} (\chi_e\ast E_y)\zt\ris\,.
\end{eqnarray}
Their solution was carried
out using a finite difference calculus
described elsewhere in detail \c{GL}. It suffices
to state here that $z$ and $t$ were discretized into
segments of size $\Delta z$ and $\Delta t$, respectively;
derivatives were replaced by central differences, and
the leapfrog method was used \c{Gers}. 

Finally,
the Fourier transform
\begin{equation}
\tilde{E}_y(z,\lambda_0) =
\int_{t_a}^{t_b}\, E_y(z,t) \exp(-i \frac{2\pi c_0}{\lambda_0}t)\, dt
\end{equation}
was calculated to determine the spectral contents of the incident and
the reflected pulses. The parameters $z$, $t_a$ and $t_b$ were
chosen to capture as much of both pulses as possible, following
the procedure described elsewhere \c{GL}. The computed ratio
of the Fourier transform of the reflected pulse to that
of the incident pulse is denoted here by $r_{TD}\om$, the
subscript TD indicating
its emergence from time--domain analysis.

\section{Numerical Results and Discussion}
For the sake of illustration, the following values
were selected for the constitutive
parameters of the dielectric--magnetic half--space: 
$p_e = 1$, $p_m=0.8$, $\lambda_e=300$~nm,
$\lambda_m=320$~nm, $M_e=M_m=100$. Thus,  $\epsri\om > 0$
and $\muri\om>0$ for all $\lambda_0$. However, $\epsrr\om$
is negative for $\lambda_0\in\les 212.1,\,300\ris$~nm, but 
it is positive
for all other $\lambda_0$; while $\murr\om$
is negative for $\lambda_0\in\les 238.6,\,320\ris$~nm, and positive
for all other $\lambda_0$. 

The definition of the refractive index in \r{rFD} suggests two
possibilities: Either 
\begin{itemize}
\item[A.] 
$\nri\om$ is negative for
 $\lambda_0\in\les 236.1,\,316.8\ris$~nm and positive elsewhere,
consistent with the requirement of $\nrr\om >0\,\forall\lambda_0$;
or
\item[B.]
$\nrr\om$ is negative for
 $\lambda_0\in\les 236.1,\,316.8\ris$~nm and positive elsewhere,
consistent with the requirement of $\nri\om \geq 0\,\forall\lambda_0$.
\end{itemize}
Thus, our attention had to be focussed on the anomalous 
spectral regime
 $\lambda_0\in\les 236.1,\,316.8\ris$~nm. In this regime, 
the reflection coefficient $r\om$
 for Possibility A is the reciprocal of that 
for Possibility B. The two possibilities can therefore
be unambiguously distinguished from one another.

The  carrier wavelength was chosen as $\lambda_{car} = 240$~nm.
The pulse duration is $3$~fs and its
3dB band is $\lambda_0\in
\les 218,\,261\ris$~nm. Therefore
the anomalous spectral regime
was substantively covered by our time--domain calculations.
The segment sizes $\Delta z = 5$~nm and $\Delta t = 0.015$~fs
used by us
were adequate for the chosen constitutive parameters, but obviously would
be totally inadequate in the resonance bands of
$\epsr\om$ and $\mur\om$.

Possibility A is clearly nonsensical.
It implies transport of energy in the half--space $z\geq z_\ell$ towards
the interface $z=z_\ell$. Not surprisingly therefore, \r{rFD}, \r{eFD} and \r{mFD} 
yielded $\vert r\om\vert >1$ for
all $\lambda_0\in\les 236.1,\,316.8\ris$~nm.

Figure 1 presents the computed values of $\vert r\om\vert$
obtained from \r{rFD}, \r{eFD} and \r{mFD} for
Possibility B (i.e., when $\nri\om \geq 0$ is guaranteed for
all $\lambda_0$). The computed values
of $\vert r_{TD}\om\vert$ are also shown therein. The 
 two sets of magnitudes compare very well for $\lambda_0\leq
290$~nm. Examination of the refracted pulse also showed
that it transported energy   away from the interface
$z=z_\ell$, which corroborates the observations
of Ziolkowski and Heyman \c{ZH}.

Thus, time--domain analysis validates the
conclusion that $\nr\om$ must be selected in frequency--domain
research in such a way
that $\nri\om\geq 0$~---~irrespective of the sign of
$\nrr\om$.

\newpage
\begin{center}
\begin{figure}[!ht]
\centering \psfull \epsfig{file=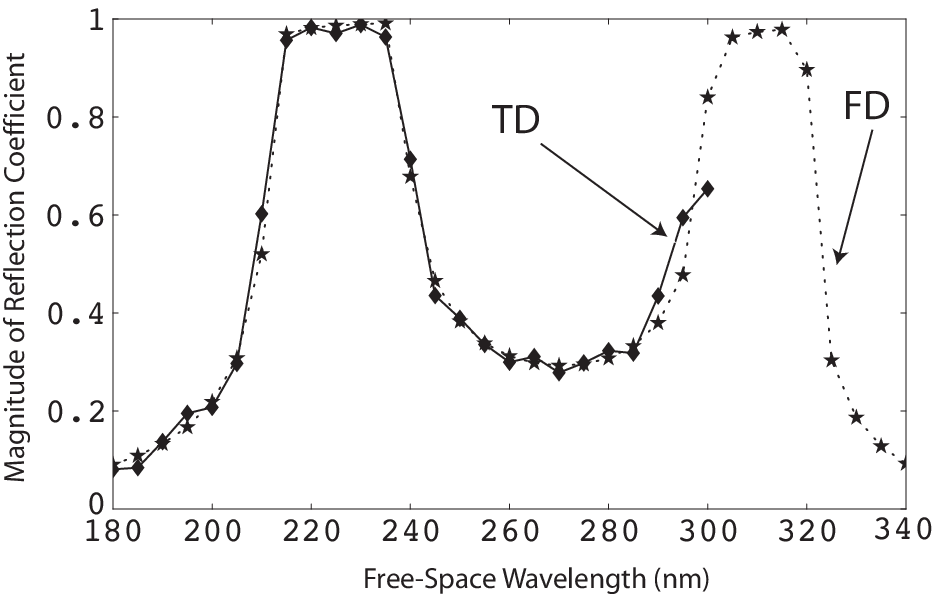,width=4in}
\caption{{\sf Magnitudes of the reflection coefficient
computed using two different methods. The frequency--domain
equations \r{rFD}, \r{eFD}
and \r{mFD}
 yield the plot
labeled FD, after ensuring that $\nri\om$ is positive for
all $\lambda_0$. The time--domain analysis, followed
by the Fourier transformations of the electric field
associated with the incident and the
reflected pulses, as discussed in Section 2, yields
the plot labeled TD. The time--domain analysis fails
in the resonance bands of $\epsr$ and $\mur$ because
the chosen discretization of space and time is neccessarily very coarse
therein.
}}
\end{figure}
\end{center}
\medskip

\end{document}